\documentclass[aps,prx,reprint,superscriptaddress,nofootinbib]{revtex4-1}

\usepackage{graphicx}
\usepackage{dcolumn}
\usepackage{bm}
\usepackage{psfrag}
\usepackage{textcomp}
\usepackage{float}
\usepackage{verbatim}
\usepackage{natbib}
\usepackage{amsmath}
\usepackage{color}
\usepackage{float}
\usepackage{comment}

\usepackage[english]{babel}
\usepackage[T1]{fontenc}

\begin{document}


\title{The Excitation Ladder of Cavity Polaritons\\}

\author{Travis M. Autry}
\email{travis.autry@colorado.edu}
\affiliation{JILA, University of Colorado \& National Institute of Standards and Technology, Boulder, Colorado 80309-0440, USA}
\affiliation{National Institute of Standards and Technology, Boulder, Colorado 80305, USA}
\affiliation{Department of Physics, University of Colorado, Boulder, Colorado 80309-0390, USA}
\author{Ga\"{e}l Nardin}
\affiliation{JILA, University of Colorado \& National Institute of Standards and Technology, Boulder, Colorado 80309-0440, USA}
\author{Christopher L. Smallwood}
\affiliation{JILA, University of Colorado \& National Institute of Standards and Technology, Boulder, Colorado 80309-0440, USA}
\affiliation{Department of Physics, University of Michigan, Ann Arbor, Michigan 48109, USA}
\affiliation{Department of Physics and Astronomy, San Jos\'{e} State University, San Jos\'{e},  California 95192, USA}
\author{Kevin Silverman}
\affiliation{National Institute of Standards and Technology, Boulder, Colorado 80305, USA}
\author{Daniele Bajoni}
\affiliation{Dipartimento di Ingegneria Industriale e dell'Informazione, Universit\'a di Pavia, via Ferrata 1 27100 Pavia, Italy}
\author{Aristide Lema\^{\i}tre}
\affiliation{Université Paris-Saclay, CNRS, Centre de Nanosciences et de Nanotechnologies, 91120, Palaiseau, France}
\author{Sophie Bouchoule}
\affiliation{Université Paris-Saclay, CNRS, Centre de Nanosciences et de Nanotechnologies, 91120, Palaiseau, France}
\author{Jacqueline Bloch}
\affiliation{Université Paris-Saclay, CNRS, Centre de Nanosciences et de Nanotechnologies, 91120, Palaiseau, France}
\author{Steven Cundiff}
\affiliation{JILA, University of Colorado \& National Institute of Standards and Technology, Boulder, Colorado 80309-0440, USA}
\affiliation{Department of Physics, University of Colorado, Boulder, Colorado 80309-0390, USA}
\affiliation{Department of Physics, University of Michigan, Ann Arbor, Michigan 48109, USA}


\date{\today}

\begin{abstract}
Multidimensional coherent spectroscopy directly unravels multiply excited states that overlap in a linear spectrum. We report multidimensional coherent optical photocurrent spectroscopy in a semiconductor polariton diode and explore the excitation ladder of cavity polaritons. We measure doubly and triply avoided crossings for pairs and triplets of exciton-polaritons, demonstrating the strong coupling between light and dressed doublet and triplet semiconductor excitations. These results demonstrate that multiply excited excitonic states strongly coupled to a microcavity can be described as two coupled quantum-anharmonic ladders.
\end{abstract}


\maketitle

The coupling of an optical absorber with an electromagnetic field is a cornerstone concept in quantum optics. When confined in a cavity, the absorber-field system can coherently exchange excitations faster than the irreversible decay rates, resulting in strong coupling as evidenced by a resolvable avoided crossing. If the zero-detuning Rabi splitting changes with the addition or loss of a single-photon, the absorber-cavity system is in the quantum strong coupling regime \cite{brune_quantum_1996,fink_climbing_2008,bishop_nonlinear_2009}. This regime is important for many applications in quantum information \cite{kimble_quantum_2008} but has only been reached when the absorber is a single two-level system \cite{thompson_observation_1992,wallraff_strong_2004,yoshie_vacuum_2004,reithmaier_strong_2004}. The use of an ensemble collectively enhances the light-matter interaction making it easier to achieve a Rabi splitting.  However, to date, the collective strong coupling regime has only realized a semi-classical Rabi splitting where the gain or loss of a single photon has little effect \cite{khitrova_nonlinear_1999}.

Despite being semi-classical, collectively coupled systems have applications in quantum networks \cite{kimble_quantum_2008} and as quantum memories \cite{simon_single-photon_2007,tanji_heralded_2009}. Additionally, in both atomic and semiconductor systems, collective strong coupling has enabled investigation of quantum phase transitions \cite{kasprzak_boseeinstein_2006,baumann_dicke_2010} and quantum fluids \cite{brennecke_cavity_2007,carusotto_quantum_2013}.

Semiconductor exciton-polaritons are a promising system to study collective strong coupling. Excitons, collective quasiparticles of bound electron-hole pairs can be confined at the antinode of a planar microcavity. The Rabi splitting of this system results in interacting quasiparticles known as the upper and lower polaritons based on their relative  energy \cite{weisbuch_observation_1992}. A rich variety of phenomena due to polariton interactions has been explored including superfluidity \cite{amo_superfluidity_2009}, bistability \cite{baas_optical_2004} and recently a weak photon blockade \cite{delteil_towards_2019,munoz-matutano_emergence_2019}. Here we utilize the polariton interactions to explore the collective excitation ladder of dressed polariton states. Dressed excitations of two-level systems have been extensively explored and described by the Jaynes-Cummings Hamiltonian,   \cite{mollow_power_1969,ulhaq_cascaded_2012,brune_quantum_1996,fink_climbing_2008,kasprzak_up_2010}. In contrast, collective systems have Hamiltonians of the form
\begin{equation}\label{eq:Ex_Polariton}
\begin{split}
H =  &\hbar\omega_{X}(k)b^{\dagger}b^{\dagger}  + \hbar\omega_{\gamma}(k) a^{\dagger}a^{\dagger} +\\  &\hbar\frac{\Omega}{2} (a^{\dagger}b + b^{\dagger}a) + H_{int}.\\ 
\end{split}
\end{equation}

\noindent This Hamiltonian dresses the exciton and $H_{int}$ includes perturbative exciton-exciton interactions  \cite{rochat_excitonic_2000}. Here $a^{\dagger}(a)$ is the photon creation(annihilation) operator, $b^{\dagger}(b)$ is the bosonic $1$s exciton creation(annihilation) operator, and $\Omega$ is the collective vacuum Rabi frequency. The strong coupling condition is realized when the Rabi frequency exceeds the loss rates of the exciton ($\gamma_{X}$) and cavity $(\gamma_{\gamma})$ i.e.   $\Omega \gg \gamma_{X} , \gamma_{\gamma}$. This Hamiltonian predicts a ladder of avoided crossings corresponding to the strong coupling of single excitons, exciton doublets, exciton triplets, etc., with light. To our knowledge, there have been no previous experimental observations of higher-order avoided crossings in any collective strong coupling system.

\begin{figure}[t!]
\centering
\includegraphics[width=3.375in]{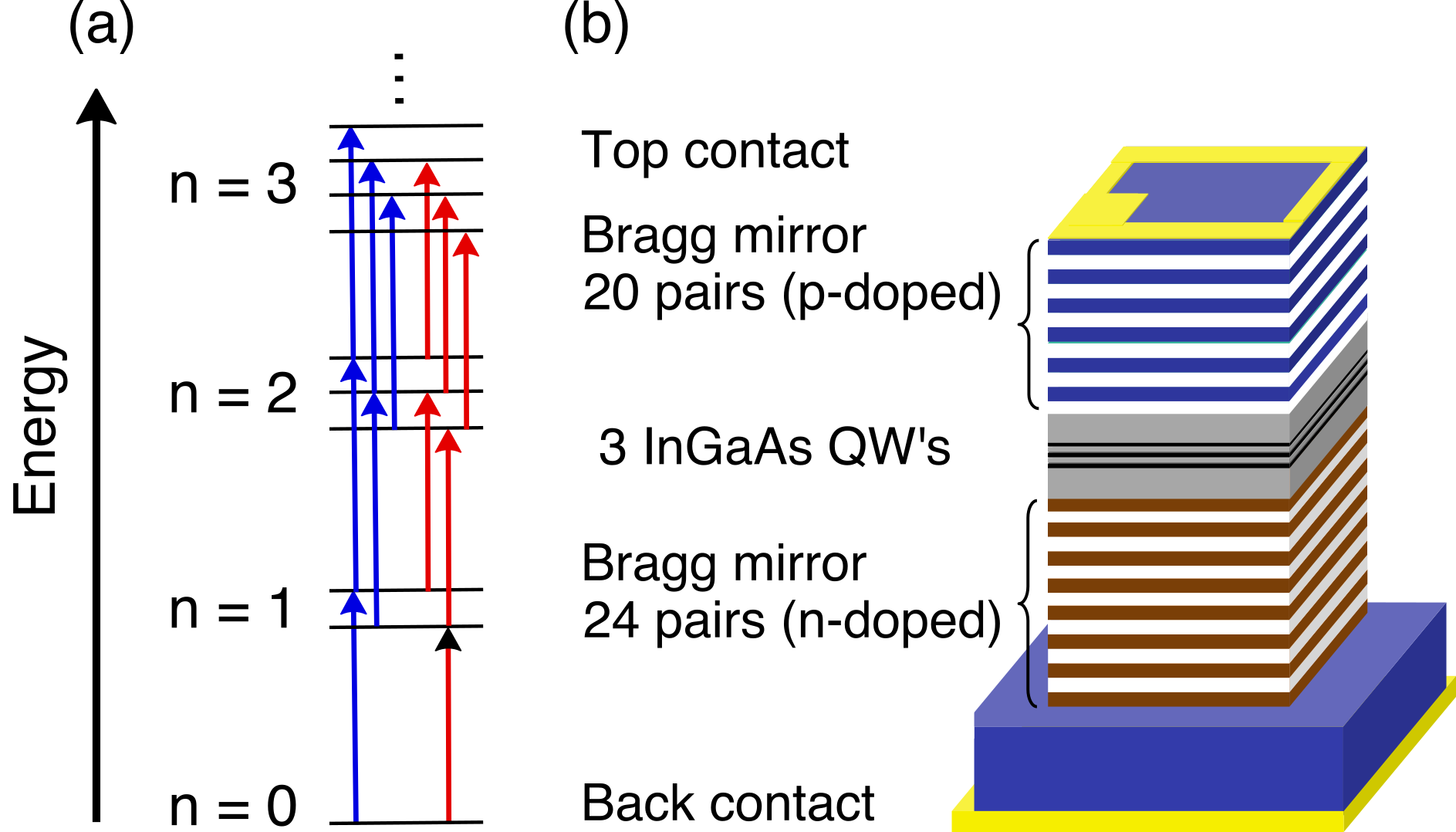}
\caption[Polariton Diode Sample] {(a) Collective strong coupling excitation ladder between $1$s excitons and cavity photons. (b) Schematic representation of the polariton diode sample. \cite{bajoni_polariton_2008}.}
\label{Diode}
\end{figure}

In this letter, we demonstrate collective strong coupling of multiply excited polariton states by measuring the avoided crossings associated with singlet, doublet, and triplet semiconductor excitations. We use a p-i-n diode GaAs microcavity to perform multi-quantum photocurrent spectroscopy and record multiply avoided crossings for the doubly and triply excited manifolds. The measured excitation ladder closely resembles that predicted by Tavis and Cummings (Fig. \ref{Diode}(a)) in the semi-classical limit of a large non-interacting absorbing ensemble strongly coupled to the electromagnetic field (coupled quantum harmonic oscillators) \cite{tavis_exact_1968,garraway_dicke_2011}.
Furthermore, the spectrum of the excitation ladder allows us to infer the dressed states of multiply excited semiconductor excitons although the detection scheme does not allow us to isolate specific excitation levels.

\begin{figure}[h!]
\centering
\includegraphics[width=3.375in]{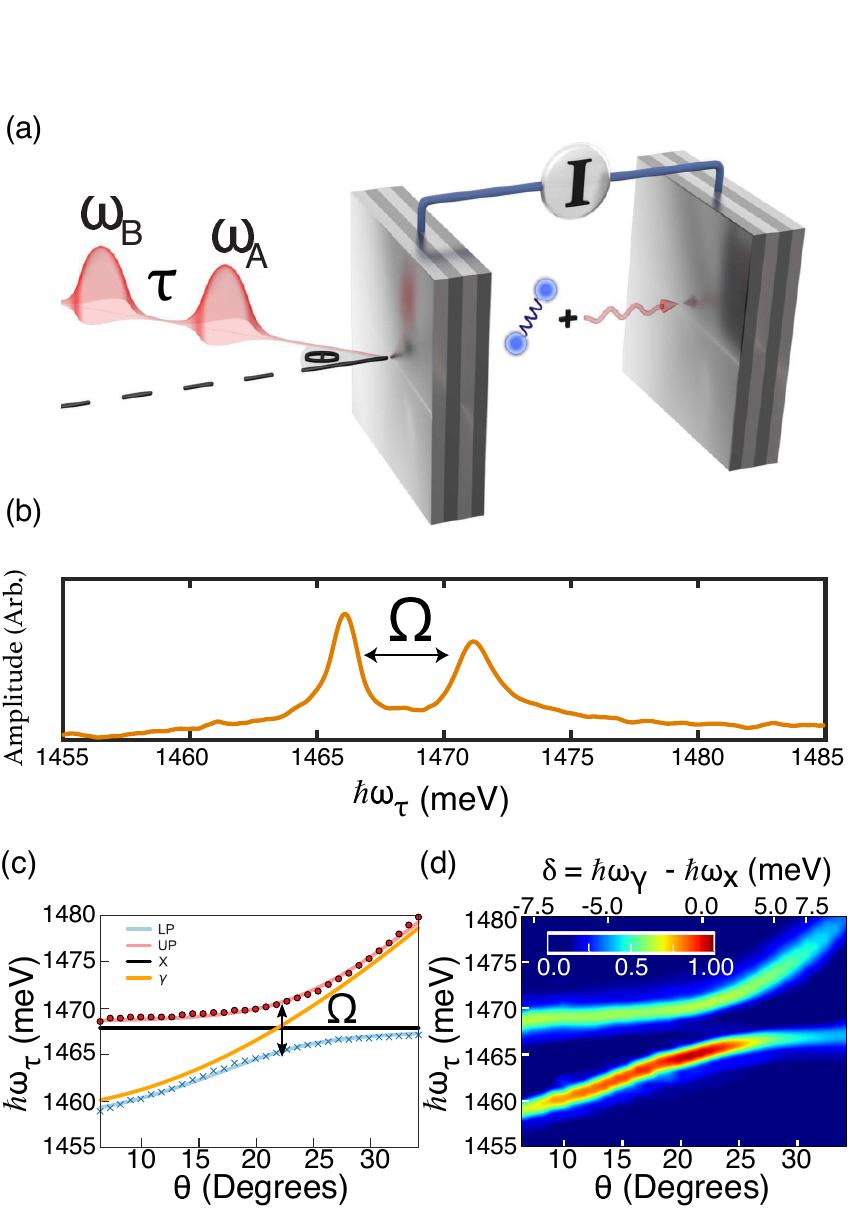}
\caption[Two-Pulse Linear Absorption Avoided Crossing] { (a) Two-pulse excitation scheme to study exciton-polaritons via photocurrent Fourier transform spectroscopy. (b) A linear spectrum taken at zero detuning shows a Rabi splitting of $\Omega \sim 5$ meV between the upper and lower polaritons. (c) Calculated dispersion curves for exciton, $X$, cavity, $\gamma$ and the strongly coupled normal modes UP and LP. The measured polariton center energies UP (dots) and LP (crosses) indicate the experimental center energies of the LP and UP spectra. (d) Contour map of linear absorption spectra as a function of detuning (top) and excitation angle (bottom). }
\label{Linear Dispersion}
\end{figure}

Measurements are performed on a microcavity diode containing three high quality InGaAs quantum wells \cite{bajoni_nonresonant_2007,bajoni_polariton_2008} (Fig. \ref{Diode}(b), details in SI) maintained at 10 K and illuminated from above. The experimental apparatus prepares four ultrafast optical pulses ($\sim$ 120 fs), each uniquely frequency shifted ($\omega_{A,B,C,D}$) by an acousto-optical modulator. Signals at the linear and nonlinear beatnotes $\omega_{lin} = \omega_{A}-\omega_B$, $\omega_{DQ} = \omega_{A}+\omega_{B}-\omega_{C}-\omega_{D}$ and $\omega_{TQ} = 2\omega_{A}+\omega_{B}-\omega_{C}-2\omega_{D}$ are recorded using a lock-in amplifier \cite{nardin_multidimensional_2013,autry_single-scan_2019}. The multi-quantum coherences associated with these signals evolve with the energy difference of an integer number of polaritons ($\Delta n =1,2, 3$). However, the selection rules for optically active transitions requires $\Delta n = 1$. Therefore, the multi-quantum measurements are sensitive to the non-radiative, $\Delta n = 2,3$, coherences, which occur through two or more $\Delta n =1$ transitions. In a manner analogous to optical-Ramsey experiments, these time-resolved coherences are mapped into an incoherent population that is measured as a modulated photocurrent \cite{nardin_multidimensional_2013}. Signals are recorded as a function of  inter-pulse delays $(\tau,T,t)$, and cavity-exciton detuning $\delta$ which is varied by changing the incident excitation angle $\theta$ (Fig. \ref{Linear Dispersion}(a)).

Recorded signals are Fourier-transformed with respect to the inter-pulse delays creating multidimensional coherent spectra (MDCS). The resulting spectrum directly shows the energy of the multiply excited states as well as any coupling. In polaritonic systems, the cavity dispersion makes polaritons sensitive to incident wavevectors. Consequently, non-collinear methods \cite{kuwata-gonokami_parametric_1997,romanelli_four_2007,kohnle_four-wave_2012,wen_influence_2013,takemura_two-dimensional_2015,wilmer_multidimensional_2015} imply a mixing of different polariton wavevectors.  Mixing different wavevectors complicates recording higher-order avoided crossings as it mixes polariton states of different excitonic fraction. In contrast, the present scheme utilizes a collinear excitation geometry that greatly simplifies the interpretation of spectra, and enables the measurement of higher-order avoided crossings.

The exciton-polariton Rabi splitting is characterized by recording linear photocurrent spectra as a function of excitation angle(detuning) $\theta(\delta)$. The zero-detuning ($\delta = \hbar\omega_{\gamma}-\hbar\omega_{x} = 0$) spectrum is shown in Fig. \ref{Linear Dispersion}(b) and exhibits two peaks split by a Rabi energy of $\hbar\Omega\sim 5$ meV.  These two peaks are the high energy upper polariton (UP) and low energy lower polariton (LP). The dressed energies and states are calculated by diagonalizing the Hamiltonian. The dressed state energies for $n=1$ are 

\begin{equation}
\hbar\omega_{UP,LP} = \frac{\hbar(w_{x}+w_{\gamma}) \pm \sqrt{\delta^{2} + \hbar^2 \Omega^2}}{2}.
\end{equation}

\noindent Plotting these energies as a function of cavity angle along with the peak amplitude positions (Fig. \ref{Linear Dispersion}(c)) provides a match to the recorded exciton-polariton dispersion curve shown in Fig. \ref{Linear Dispersion}(d). At zero detuning, the dressed states corresponding to these energies are equally weighted superpositions between exciton and cavity:

\begin{equation}\label{eq:UP}
|\psi_{UP,LP}\rangle = \frac{|1_{x},0_{\gamma}\rangle \pm |0_{x},1_{\gamma}\rangle}{\sqrt{2}} .
\end{equation}

\begin{figure}[h!]
\centering
\includegraphics[width=3.25in]{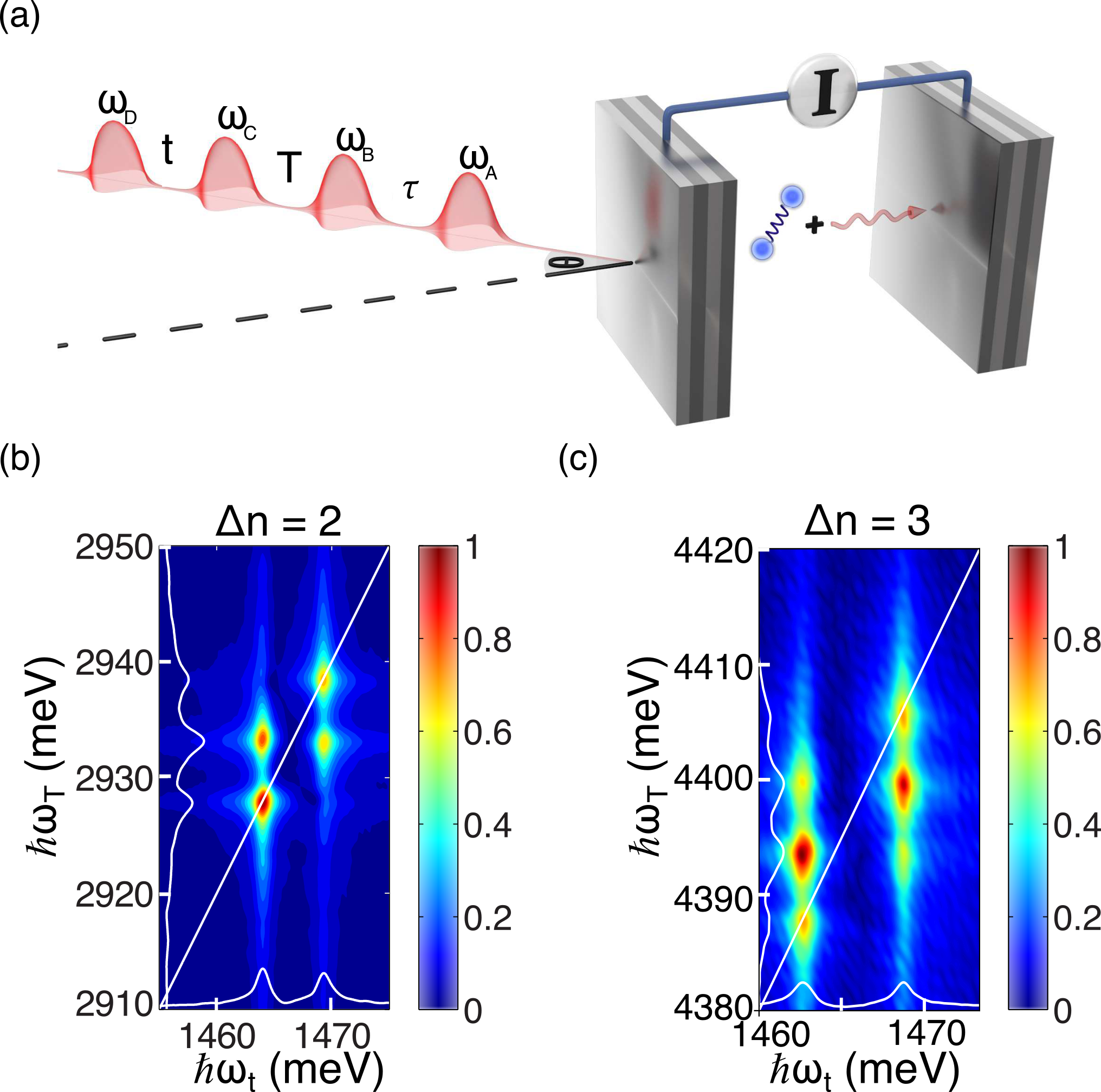}
\caption[Four-Pulse MDCS Experiment and Spectra] { (a) Four pulse excitation scheme to study exciton-polaritons via photocurrent MDCS. (b) The magnitude of a two-quantum MDCS spectrum ($\omega_{DQ}$). (c) The magnitude of a three-quantum MDCS spectrum ($\omega_{DQ}$).}
\label{Four-Pulse MDCS}
\end{figure}

To measure polariton pairs and triplets, multidimensional spectra are recorded (Fig. \ref{Four-Pulse MDCS}(a)) at the nonlinear beatnotes $\omega_{DQ},\omega_{TQ}$. The signals are recorded as a function of angle and the inter-pulse time delays \textit{T} and \textit{t} with $\tau=0$. The multi-quantum coherence evolves during the time delay \textit{T}. All spectra are normalized to $1$ due to different experimental conditions (discussed in the SI) for the different pulse sequences. The double-quantum spectrum (Fig. \ref{Four-Pulse MDCS}(b)) reveals four peaks split by the Rabi frequency $\Omega$. The two diagonal peaks correspond to pairs of lower or upper polaritons. The two off-diagonal peaks represent mixed pairs of lower and upper polaritons. The projection (Fig. \ref{Four-Pulse MDCS}(c)) onto the $\hbar\omega_{t}$ axis shows a conventional polariton spectrum where the two mode frequencies UP and LP correspond to $\Delta n=1$ coherences. The projection onto the $\hbar\omega_{T}$ axis shows the frequencies corresponding to $\Delta n=2$ coherences.

Additional measurements recording three quantum coherences, $\Delta n=3$, (detected at $\omega_{TQ}$) reveal polariton triplets.  The spectrum resulting from these triplets (Fig. \ref{Four-Pulse MDCS}(c)) reveals six peaks with two peaks along the diagonal. The four off-diagonal peaks, like the off-diagonal peaks of the two-quantum coherence, represent mixed states with quantum pathways capable of radiating with the energy of either an upper or lower polariton.  

\begin{figure}[h!]
\centering
\includegraphics[width=3.375in]{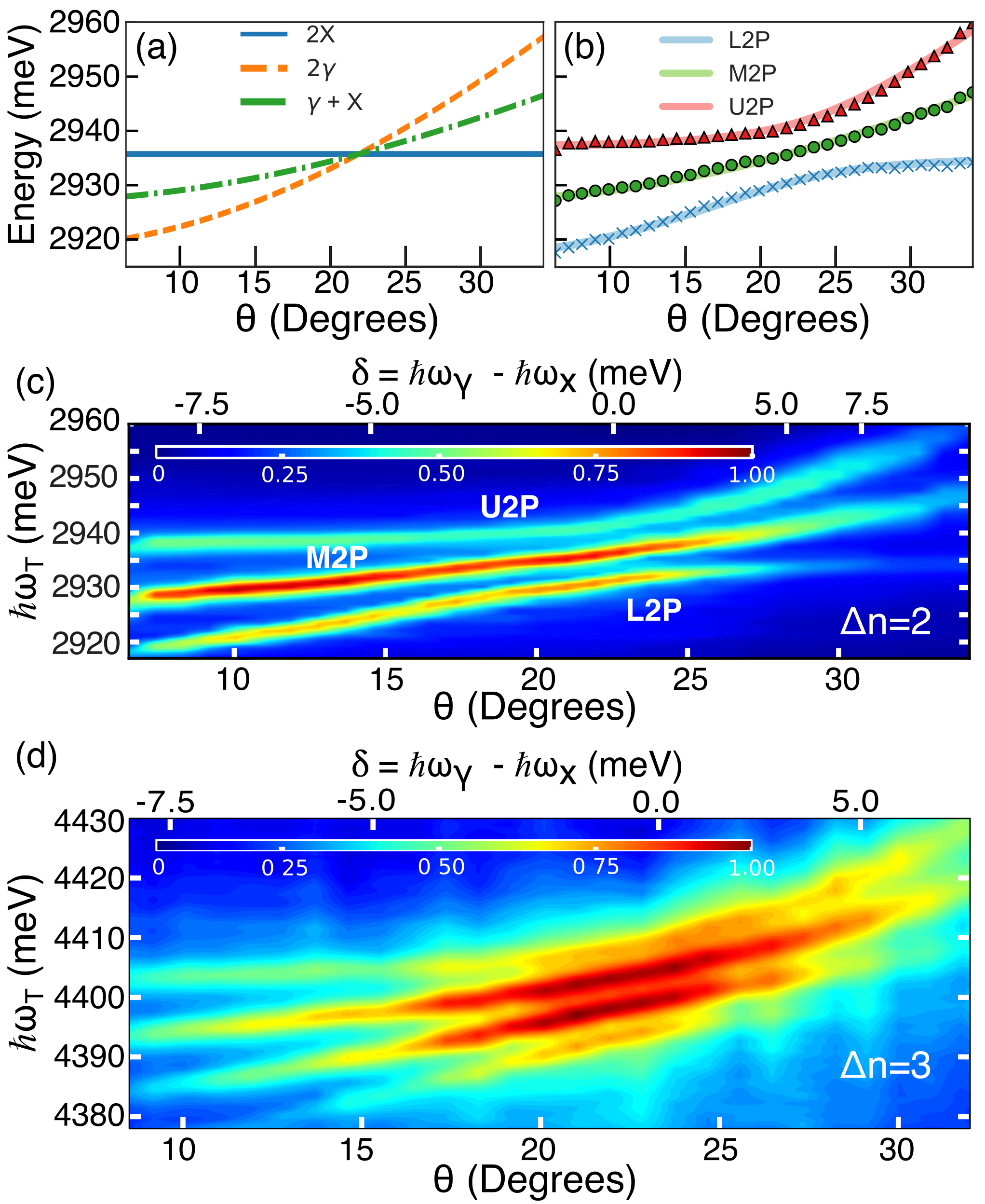}
\caption[Higher Order $\Delta n =2,3$ Avoided Crossings for Exciton-Polaritons] { (a) Calculated dispersion of a doubly excited exciton state ($2X$), a doubly excited cavity ($2\gamma$) state, and a ($X + \gamma$) state.  At the avoided crossing angle $\theta_{0}$ all three energies are degenerate. (b) Calculated strongly coupled exciton and cavity pairs as a function of excitation angle. At $\theta_{0}$ a double avoided crossing is calculated. The measured polariton-pair center energies U2P ($\bigtriangleup$), M2P ($\bigcirc$), and L2P (X) are shown for reference. (c) Integrated double-quantum spectra as a function of excitation angle-detuning. (d) Integrated triple-quantum spectra as a function of angle-detuning.}
\label{2Q_Dispersion}
\end{figure}

We calculate higher order dispersion curves (Fig. \ref{2Q_Dispersion}(a-b)) to show the uncoupled and coupled $\Delta n=2$ cavity-exciton modes from Eq. \ref{eq:Ex_Polariton} as a function of excitation angle $\theta$. The dressed state energies of doubly excited polaritons are

\begin{equation}\label{eq:polariton_energy}
\hbar\omega_{U2P,L2P} =  \hbar( w_{x}+w_{\gamma}) \pm \sqrt{\delta^{2} + \hbar^2 \Omega^{2}},
\end{equation}

\begin{equation}\label{eq:polariton_energy2}
\hbar\omega_{M2P} = \hbar (w_{\gamma} + w_{x}).
\end{equation}

Interestingly, these energies predict a doubly-avoided crossing with a Rabi splitting, $\Omega$, of $\sim5$ meV at zero-detuning as shown in Fig. \ref{2Q_Dispersion}(b). At zero detuning, the dressed $n=2$ states are

\begin{equation}\label{eq:U2P}
|\psi_{U2P,L2P}\rangle = \frac{|2_{x},0_{\gamma}\rangle + |0_{x},2_{\gamma}\rangle \pm \sqrt{2}|1_{x},1_{\gamma}\rangle}{2}, 
\end{equation}

\begin{equation}\label{eq:M2P}
|\psi_{M2P}\rangle = \frac{|2_{x},0_{\gamma}\rangle - |0_{x},2_{\gamma}\rangle}{\sqrt{2}}.
\end{equation}

These higher order avoided crossings are measured by integrating the MDCS spectra along $\hbar\omega_{\textit{t}}$ and plotting the frequency integrated double-quantum spectrum as a function of excitation angle (Fig. \ref{2Q_Dispersion}(c)). The measurement matches the calculations exhibiting two avoided crossings at zero-detuning $\sim 22$\textdegree. The observed spectrum corresponds to pairs of polaritons labeled $U2P$ for two upper polaritons, $L2P$ for two lower polaritons and $M2P$ for a mixture of upper and lower polariton. We also calculate the expected spectrum and dispersion (not shown) for $\Delta n=3$ coherences and find that we expect three avoided crossings. The experimental spectra (Fig. \ref{2Q_Dispersion}(d)) indeed exhibits a triply avoided crossing between four modes each with a Rabi splitting of $\sim 5$ meV.

These measurements demonstrate strong coupling of multiply excited polariton states. Using the few-particle description, these measurements are consistent with interactions providing only a perturbative correction to the optical response of exciton-polaritons. This conclusion is realized by calculating the eigenstates and eigenenergies of $H$ during inter-pulse delay $T$ while treating the interactions as a perturbation on the polariton energies and linewidths. However, the absence of interaction terms $H_{int}$  would cause a zero amplitude signal due to complete destructive interference for all time-delays $t$. Because the amplitude of the measured signals requires anharmonicity \cite{baum_multiplequantum_1985,ernst_principles_1990,yang_two-dimensional_2007,yang_revealing_2008,stone_two-quantum_2009,karaiskaj_two-quantum_2010}, this pulse sequence bypasses the usual requirement to probe the excitation ladder through energy shifts or changes in the Rabi splitting, which may be partially or totally unresolved as in the case of dilute atomic systems where the interactions are a small effect \cite{gripp_anharmonicity_1996,thompson_nonlinear_1998}.  Therefore, the interactions introduce sufficient anharmonicity in the coupled excitation ladder (Fig. \ref{Diode}(a)) to observe these signals but are not sufficiently large enough to modify the excitation ladder, as in the case of quantum strong coupling or polariton blockade.

Our calculations show that the measured spectra agrees with the excitation ladder expected from a strongly coupled non-interacting collective ensemble \cite{tavis_exact_1968,garraway_dicke_2011}. Intuitively, because a collective ensemble has a larger phase-space, it supports more excitations than a single two-level system. Thus, the eigenbasis of a large collective ensemble (excitons in semiconductor/molecular physics, Dicke states in atomic physics) is well described by a bosonic ladder of states and bosonic descriptions of these systems have been developed using the Usui (Holstein-Primakoff) transformations \cite{rochat_excitonic_2000,garraway_dicke_2011}. Therefore for weak excitation, strongly coupled collective systems have the semi-classical eigenstructure expected for two coupled quantum harmonic oscillators \cite{garraway_dicke_2011} as observed in Fig. \ref{2Q_Dispersion}. This excitation ladder differs from previously observed dressed states of two-level systems \cite{mollow_power_1969,brune_quantum_1996,fink_climbing_2008,bishop_nonlinear_2009,ulhaq_cascaded_2012} and an interesting result of these calculations is that while the $n=1$ dressed states are the same as the solutions to the Jaynes-Cummings Hamiltonian for $n=1$, the $n=2,3...m$ solutions are not the same (see for instance eigenstates in equations (\ref{eq:polariton_energy}-\ref{eq:M2P})). These dressed states may be a resource for future quantum optical applications.

The optical properties of strongly coupled systems can be intuitively discussed from the viewpoint of their respective excitation ladders. Strong anharmonicity is introduced in Jaynes-Cummings systems because only a single-excitation is supported by the two-level system. This leads to an anharmonic-coupled excitation ladder that exhibits a changing Rabi splitting. In contrast, collective systems that support many excitations, such as exciton-polaritons, achieve anharmoncitity through many-body interactions \cite{delteil_towards_2019,munoz-matutano_emergence_2019}. As a result, enhanced interactions in novel devices and configurations have been engineered to achieve quantum optical effects  \cite{jia_strongly_2018,rosenberg_strongly_2018,togan_enhanced_2018,knuppel_nonlinear_2019}. 

\begin{figure}[b]
\centering
\includegraphics[width=3.25in]{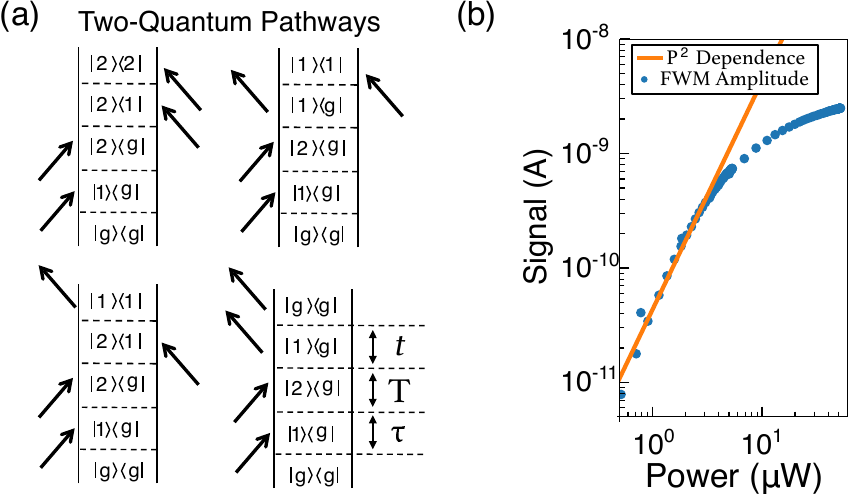}
\caption[Diagrams and Power.] { (a) Schematic showing the quantum pathways excited through a double-quantum pulse sequence. (b) Power dependence taken at zero-detuning recorded for the two-quantum ($\Delta n=2$) signals showing the signal amplitude relative to total excitation power.}
\label{3Q_Power}
\end{figure}

Coherent nonlinear spectroscopy allows particle pairs to be studied \cite{yang_two-dimensional_2007,yang_revealing_2008}. This can be understood from excitation pathways as described by double-sided Feynman diagrams. The double-quantum diagrams (Fig. \ref{3Q_Power}(a)) illustrate polariton pair excitation paths. 
In the present notation the diagrams describe coherences between the ground, first, and second excitation manifolds ($g,1,2$), with the diagrams for the individual states presented in the SI. These excitation pathways come in pairs with each term nearly identical but opposite in sign. By recording at a unique frequency, only processes belonging to the sum of these terms are measured. In the absence of interactions, these terms are identical, resulting in perfect destructive interference (zero-amplitude) \cite{yang_two-dimensional_2007,yang_revealing_2008} These excitation pathways are valid if a second-order power dependence of the four-wave-mixing signal is observed (Fig. \ref{3Q_Power}(b)). This power dependence is expected for the few particle picture in multidimensional spectroscopy. 

In conclusion, our measurements explore the exciton-polariton ladder of states demonstrating multiply avoided crossings associated with the dressed states of multiply excited excitons. The higher order avoided crossings exhibits the characteristic semi-classical spectrum of two coupled quantum anharmonic oscillators with anharmonicity due to polariton-polariton interactions. Interactions are strong enough for the excitation ladder to be detected by non-linear spectroscopy, but remain weak enough to not cause any quantum optical deviation to the collective strong coupling ladder. The spectra are reproduced using a dressed exciton picture. These measurements establish exciton-polaritons as a platform to explore collective strong coupling and we expect that this technique can be adapted to other collective systems exhibiting interactions such as Rydberg atoms and trapped ions.

\acknowledgements
We thank James Thompson and Bo Sun for helpful discussions. The research leading to these results was funded by NSF grant (1415398), the JILA PFC (1125844), and by the French RENATECH network.

\bibliography{ref_v14}

\end{document}